\documentclass[conference]{IEEEtran}
\usepackage{graphicx}
\usepackage[noadjust]{cite}
\usepackage[acronym]{glossaries} 
\usepackage{lipsum}
\usepackage{fancyhdr}
\begin{document}

\title{Building Airflow Monitoring and Control using Wireless Sensor Networks for Smart Grid Application}

\author{\IEEEauthorblockN{Nacer Khalil}
\IEEEauthorblockA{Computer Science Department\\
University of Houston}
\and
\IEEEauthorblockN{Driss Benhaddou}
\IEEEauthorblockA{Engineering Technology Department \\
University of Houston}
\and
\IEEEauthorblockN{Abdelhak Bensaoula}
\IEEEauthorblockA{Physics department \\
University of Houston}
\and
\IEEEauthorblockN{Michael Burriello}
\IEEEauthorblockA{Facilities Management \\
University of Houston}

\and
\IEEEauthorblockN{Raymond E Cline, Jr.}
\IEEEauthorblockA{Information and logistics Technology Department \\
University of Houston}
}
\maketitle
\thispagestyle{fancy}

\begin{abstract}
The electricity grid is crucial to our lives. Households and institutions count on it. In recent years, the sources of energy have become less and less available and they are driving the price of electricity higher and higher. It has been estimated that 40\% of power is spent in residential and institutional buildings. Most of this power is absorbed by space cooling and heating. In modern buildings, the HVAC (heating, ventilation, and air conditioning) system is centralised and operated by a department usually called the central plant. The central plant produces chilled water and steam that is then consumed by the building AHUs (Air Handling Units) to maintain the buildings at a comfortable temperature. However, the heating and cooling model does not take into account human occupancy. The AHU within the building distributes air according to the design parameters of the building ignoring the occupancy. As a matter of fact, there is a potential for optimization lowering consumption to utilize energy efficiently and also to be able to adapt to the changing cost of energy in a micro-grid environment. This system makes it possible to reduce the consumption when needed minimizing impact on the consumer. In this study, we will show, through a set of studies conducted at the University of Houston, that there is a potential for energy conservation and efficiency in both the buildings and the central plant. We also present a strategy that can be undertaken to meet this goal. This strategy, airflow monitoring and control, is tested in a software simulation and the results are presented. This system enables the user to control and monitor the temperature in the individual rooms according the locals needs. 
\end{abstract}
\section{Introduction}
Energy scarcity is increasingly becoming an important challenge that nations are faced to solve. In fact, 40\% of the energy is spent in buildings \cite{ref1}, and a significant amount is used for space cooling/heating. This is the case for the University of Houston, located in Houston, Texas, known for being a very warm and humid city. As a result, it spends a considerable amount of money to maintain a convenient temperature and humidity level. It counts on modern and highly efficient HVAC (heating, ventilation, and air conditioning) systems to maintain the buildings at a convenient temperature and humidity level. Many of the rooms in every building are cooled without being used which represents a loss. Some of the buildings are rarely used while others are used very differently throughout the day (e.g. labs are many times empty and other times full). This results in considerable inefficiency when maintaining different rooms equally.\par
The way the current air-handling units work is that it ties all rooms with constant quality of cooled air. This keeps them at the same temperature, but some of the air is removed even if it is cool enough to be kept. In addition, there is no way to customize the temperature according to one's needs. Usually, one cannot change the rooms' temperature without affecting the temperature of the other rooms on the same floor. This is a major drawback of the current air handling unit in addition to loss it creates when not accounting of the occupancy of the rooms. \par
The drawbacks of the  current system are an important motivation for our team to build an overlying system that solves these drawbacks by increasing users' comfort and decreasing the energy loss. This system uses a Wireless Sensor Network (WSN) to sense the temperature and to control the quantity of air flowing into every room.\par
The main challenge was to design the air flow management system without replacing the current system. We propose a system that uses WSN where each sensor node is placed in room of the building and another one placed in the air distribution system. This makes it possible to sense in real-time the temperature and then, choosing the appropriate angle to open the damper through which air will flow into the room based on a model. The model will be presented, in addition to a simulation that was conducted to test the system's ability to serve the user and increase energy efficiency. \par
The rest of the paper is organized as follows: Section II presents the background and related. In section III, the methodology is presented. In section IV, the results of the simulation are presented and discussed. Section V concludes the paper and presents the future work.

\section{Background and related work}
Energy is one of the most important concerns of the world. As fossil fuels become less and less available, their price has increased significantly in the last decade, and raises concerns for future generations' energy outlook. Electricity is mostly generated using coal, which has a huge negative impact on the environment \cite{ref7}. \par
In recent years, due to high electricity cost, high growth rate of the electric grid and climate change, the one century-old electric grid has seen a drastic shift in its internal design and philosophy \cite{ref8}. The smart grid is intended to support Distributed Energy Generation (DER), two-way power and communication flow between the producer and consumers of power, and increased usage of renewable energies \cite{ref9}. This impacts the price of electricity where it becomes variable and changes very frequently \cite{ref10}. The user can benefit by minimizing the usage when the price is high and shift the necessary tasks to the times when the price is low. However, the user cannot do this manually as it is very time consuming. Smart buildings have the potential to automate the process in addition to minimizing the losses in a building \cite{ref11}. \par
In an industrial environment, the buildings count on the services delivered by the central plant for providing electric power with high reliability for cooling and heating. The central plant is an efficient way to manage power, because it uses the economy of scale of larger machinery that is more efficient to serve multiple buildings rather than smaller machinery that is neither as efficient nor as reliable \cite{ref12}. \par
The buildings are the main consumers of electric power, chilled water and steam. They use these three sources to power the users' machinery and provide them with a comfortable space for living \cite{ref13}. The buildings can be optimized to avoid energy loss\cite{ref14}. \par
The buildings as well as the central plant can become more efficient using more automation that takes into consideration the electricity price, human activity and machines efficiency profiles. In \cite{ref15}, they use sensing to monitor the building's consumption as well as the users' occupancy and incorporate this data into agents to estimate the occupants in the building, predict them and adjust the building's consumption to this changing environment. In \cite{ref16}, sensory data is collected every fifteen minutes in order to determine the consumption of the building in the next hour. \par
In modern buildings, there are Air Handling Units that consume chilled water and steam to adjust the environment for the occupant. There are different types of Air Handling Units, such as fan Coils, which are small and simple and are commonly used in smaller buildings and commercial applications. On the other hand, there are custom AHUs available in any configuration that a user might require, which are very common for institutional and industrial applications. \par
There are  a number of research activities towards saving the energy consumption in Buildings. One of those is an energy efficient model predicting building temperature control \cite{ref3}. This work focuses on the problem of controlling the vapor compression cycle (VCC) in an air-conditioning system, containing refrigerant which is used to provide cooling. Another paper on Experimental investigation of wraparound loop heat pipe heat exchanger used in energy efficient air handling units \cite{ref4}. This is an experimental investigation on the thermal performance of an air-to-air heat exchanger, which utilises heat pipe technology. While most of the research is in the area of HVAC (Heating, Ventilation and Air Conditioning) our model deals with only air flow. \par
\section{Data gathering and analysis}
In this section, we show the different studies that were undertaken to investigate the potential for energy optimization in the central plant and the buildings. The gathered data for the central plant is a collection of data logs that monitor the status of every machine of the central plant every two hours. We used four months of data for this study.
\subsection{Central plant optimization}
\begin{figure}[htbp]
\centering
\includegraphics[scale=0.35]{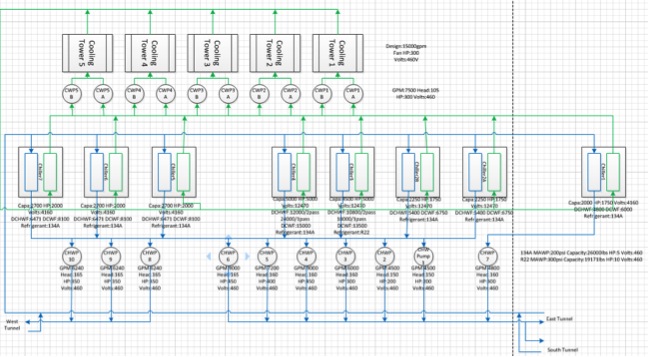}
\caption{University of Houston Central plant machines map}
\label{cp}
\end{figure}

The central plant is composed of a set of machinery that provide chilled water and steam. These two systems are independent as they do not share any machinery. However, each of these systems is composed of a set of machines that work hand in hand to provide the required service.
To develop the Airflow Management System using Sensor Networks, different parameters had to be taken into account to design the system which enables a user to control the air temperature of the room. To understand the scenario, let us consider the schematic in Figure \ref{cp}. The main types of machines that are used are: chillers, chilled water pumps, condenser water pumps and cooling towers. For each type, there are a number of machines consisting of different brands, different ages and different usage hours. As a result, the efficiency profile of every machine is different and thus affects uniquely the efficiency of the whole system. \par
\begin{figure}[htbp]
\centering
\includegraphics[scale=0.35]{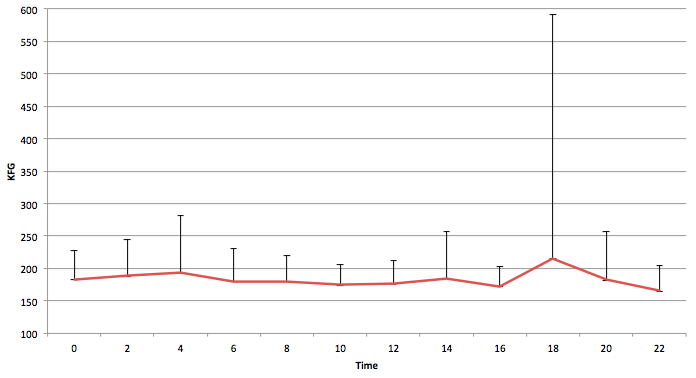}
\caption{Evolution of KFG by Hour}
\label{fig1}
\end{figure}

To evaluate the efficiency of the chiller system, we used the following measure called KFG, which stands for the number of KiloWatt Hour required to cool 10,000 GPM of chiller by 1 degree Fahrenheit (F) in one hour. Figure \ref{fig1} shows that the efficiency of the system varies throughout the day and therefore there is potential to investigate more to understand the driving factors of such variation in cost. \par
To do so, we defined two additional measures. The first variable, $\Delta T$, refers to the temperature difference between the supply chilled water and the return chilled water temperature in degrees Fahrenheit. The second variable, NT, is equal to AHU chilled water set point minus the return temperature of the chilled water. The set point for the AHU is equal to 55.1F. This measure gives an indication of how much chilled water was consumed by the buildings to estimate the overproduction of chilled water.
\begin{figure}[htbp]
\centering
\includegraphics[scale=0.4]{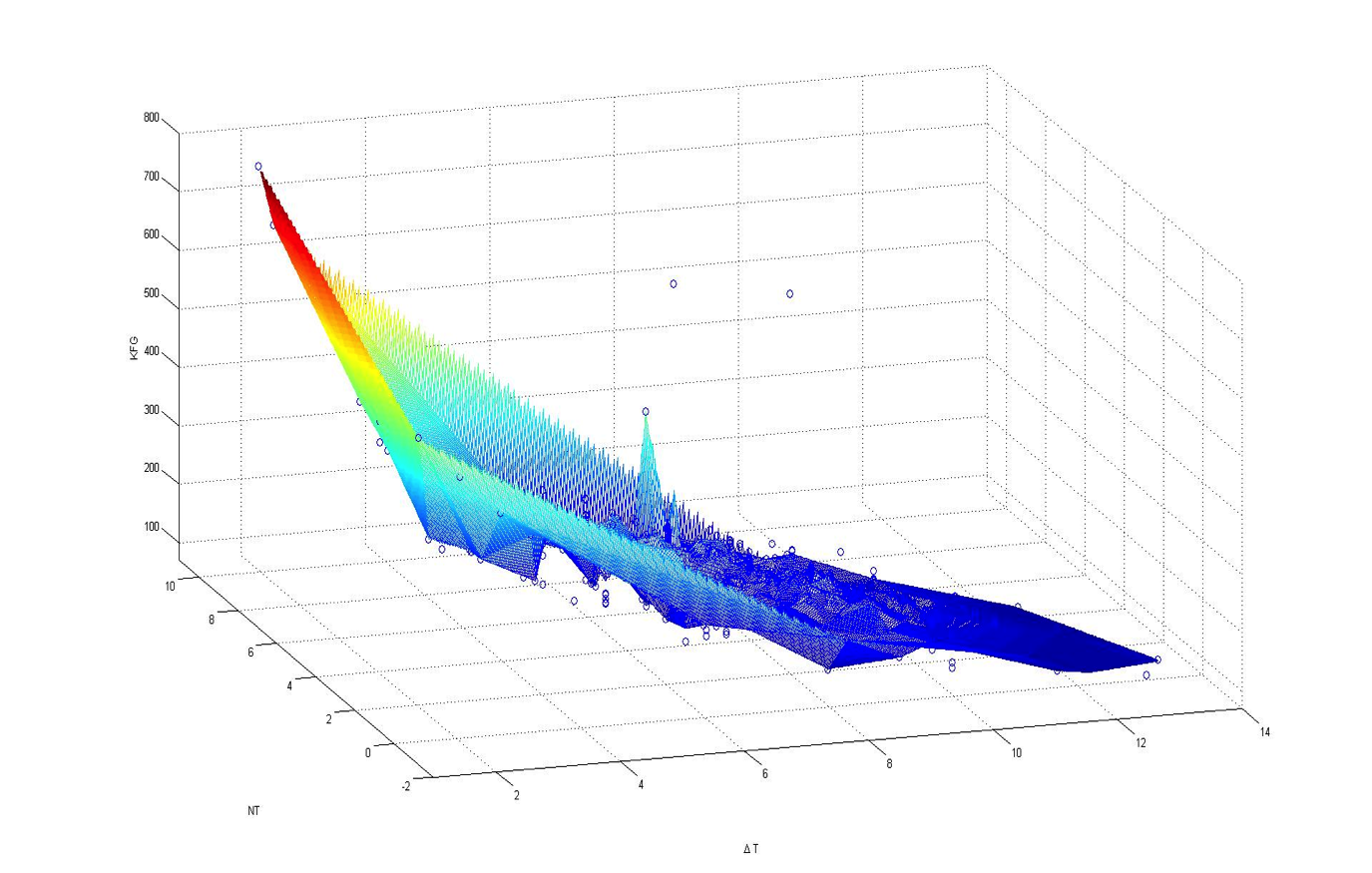}
\caption{Evolution of KFG as a factor of NT and $\Delta T$}
\label{fig2}
\end{figure}

From Figure \ref{fig2}, it shows that KFG is affected by $\Delta T$ and NT. This means that the more the heat is used for the same quantity of water, the more efficient it is. Also, the closer the chilled water return temperature is, the more efficient it is.

\subsection{Building Optimization}
The buildings are the main consumers of chilled water and steam, they use them to cool/heat the air and to remove humidity. There are many parameters that affect the production of air. A study was conducted to estimate the variables that affect the consumption. The chosen parameters are the state of doors, human presence and machines used. \par
\begin{figure}[htbp]
\centering
\includegraphics[scale=0.37]{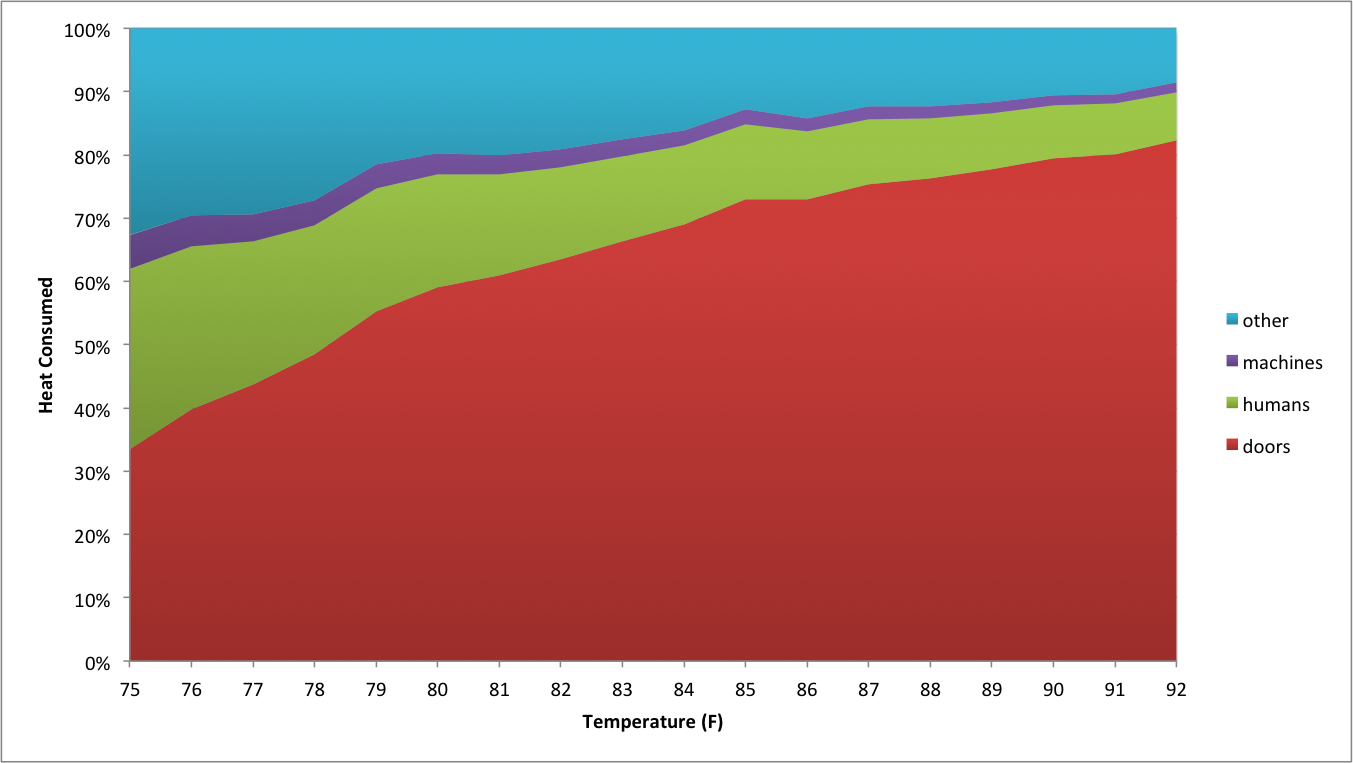}
\caption{Heat consumption in a building}
\label{fig3}
\end{figure}
Figure \ref{fig3} shows how these variables consume heat when the outside temperature  varies. It is clear that the doors' states represent the major consumers of heat in a building. The fact that opening a door for a few seconds results in an important heat exchange between the building and the outside air. This opening of doors is caused by human activity. Therefore, there is a potential for energy minimization by modeling and predicting human activity and its interaction with the building. This model can be incorporated with an airflow monitoring and control system to minimize the losses caused by opening doors.

\section{Simulation}
The goal of this simulation is to monitor and control the airflow speed which is the main consumer of chilled water and steam. By optimizing the airflow, we optmimize the consumption of chilled water and steam and this affects not only the buildings but also the central plant efficiency. The following simulation is used to study the airflow system in a typical room. It adjusts to the user's requested temperature by taking into account his/her request and therefore makes the room more comfortable while consuming less energy.
\subsection{Scenario}
When the user feels that it is too warm for him/her, the user requests the system to change target temperature and specifies how long the system has to achieve it. By getting these two parameters, the system adjusts the damper angle to allow more/less air to flow to meet the target temperature by the required time.
\subsection{System requirements}
The system has a set of requirements that drive its architecture and design, as per the following
\begin{itemize}
\item Achieve target temperature within the required time
\item Keep that temperature constant unless changed
\item Account for change in occupancy that will impact achieving the target within allocate time
\end{itemize}
The model is designed to take into consideration that once a quantity of air is determined, the system should always recompute it when the occupancy of the room changes as more people can come/leave the room, adding/removing more heat and therefore deviate the system from its objective.\par
The system therefore works as follows: Once the target temperature and time are received from the user, the system senses the temperature of the room, the temperature of the air flowing to the room, computes the angle of the damper using the model to compute the next minute's partial objective and adjusts the shed to that angle. After each time period, the temperature is expected to be at a certain level, so the system senses the temperature of the room to see if the partial objective has been met. Afterwards, it recomputes the angle using the model to take into consideration the error arising from the previous partial objective.
\subsection{Model}
The model counts on the following assumptions and parameters some of which are set as constant in the system  and do not need to be sensed and others that change and must be sensed every time step in order for the system to achieve the objective, as follows \par
\subsubsection{Assumptions}
The model assumes some things are constant to keep it simple. The system assumes the pressure inside the room is the same for every room and does not change throughout time. Also, it assumes that the speed of air, $v_{out}$ is the same as the speed of the air flowing to the room. In reality, the direction of the air flow vector affects the speed. To keep it simple, we assume these parameters.

\subsubsection{Parameters}
\begin{itemize}
\item Target temperature $K_{target}$: This is the input temperature provided by the user. It is the target temperature to be achieved by the system.
\item Time $\Delta t_{target}$: This is the input time in minutes provided by the user. It is the time required to meet the target temperature.
\item Output $\alpha$: This is the angle of the damper that is computed by the model. It is what defines the quantity of air flowing to the system.
\item Inside temperature $K_{in}$: This is the inside room temperature. It is sensed by the sensor node within the room. It is sensed every minute. It is also used to check the error in meeting the partial target.
\item Outside Air temperature $K_{out}$: This is the temperature of the air in the duct and it is sensed every minute because it can be changed by the air-handling unit and it taken into account to compute $\alpha$.
\item Outside air Speed $v_{out}$: This is the speed of the air flowing in the duct.
\item Volume of the room: $V_{room}$: This parameter represents the volume of the room.
\item Air density $\rho = 1.225 Kg/m^3$
\item Surface of damper $l^2$: This is the surface of the damper that we change. The bigger the $\alpha$, the bigger the surface, the more air flows into the room.
\end{itemize}
The model we have created for this study outputs the angle $\alpha$ that is computed using the parameters above. \\ \\
$\alpha = tan^{-1} (\frac{\rho *V_{room}*(K_{target}-K_{in})}{\Delta t_{target}* v_{out}*L^2*(K_{out}-K_{in})})$ \\ \\
This model computes the angle of the damper to ensure enough air is flowing to the room to meet the target temperature within the required time. It is based on thermodynamic laws and most importantly the law of conversation of energy.
\section{Results}
The purpose of this section is to test whether the model will make it possible to reach the goal of customizing the temperature to the users' needs and be able to increase both user's comfort and energy efficiency. To do so, a program was built that simulates the change of temperature by adding and removing people. It also simulates internal room heat that changes with time (e.g. computers, heat from windows...). Afterwards, the model was added to the program to see if it can successfully meet the user's target within the target time.
\subsection{Simulation}
The simulation program is a Java SE \cite{Java7Spec} program that depicts a room with users going in and out of the room and controls the air damper to meet the target temperature within the target time.
\subsection{Data collection}
The simulation is run assuming the following information. The room size is 150 cubic meters. The air speed is 100 meters per minute. The starting temperature is 24 degrees celsius. The user asks the system to achieve a temperature of 21 degrees Celsius within 30 minutes. In addition, people go into the room and exit randomly. This generates heat randomly. The system takes into consideration such possible changes to adjust to in order to meet the target.
\begin{figure}[htbp]
\centering
\includegraphics[scale=0.21]{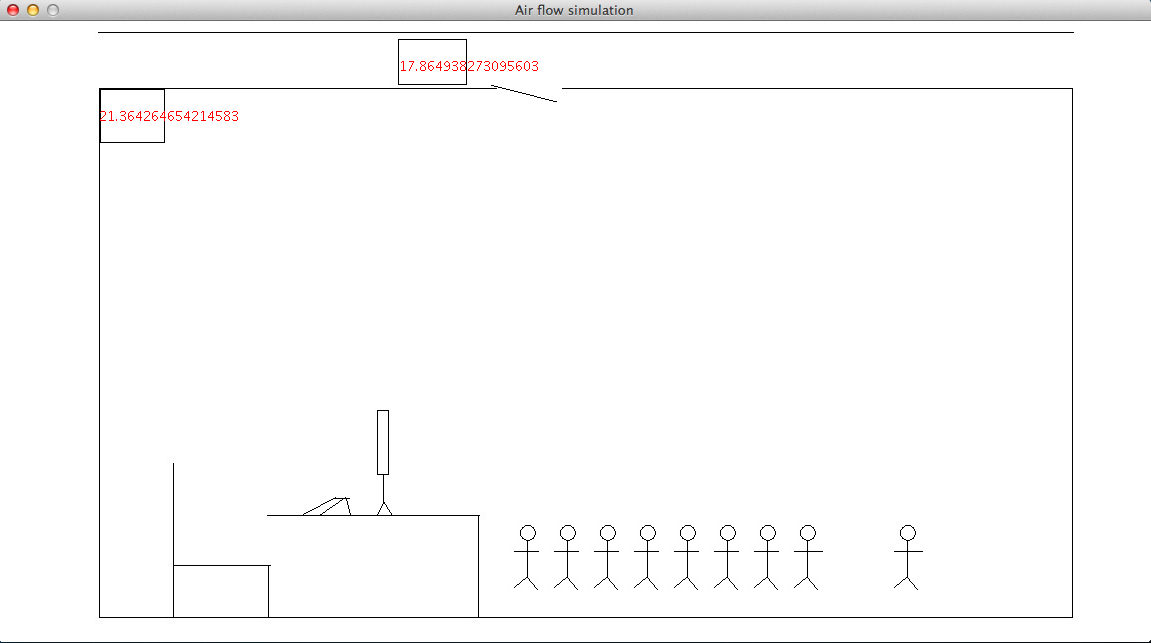}
\caption{Java airflow simulation program}
\label{figx}
Figure \ref{figx} shows a screenshot of the Java simulation application that simulates the room and shows the model in action. The simulator outputs after every minute the current temperature of the room and the angle of the damper that will be set to meet the target given the time left. This data was collected to evaluate the performance of the system.
\end{figure}
\subsection{Data interpretation}
\begin{figure}[htbp]
\centering
\includegraphics[scale=0.52]{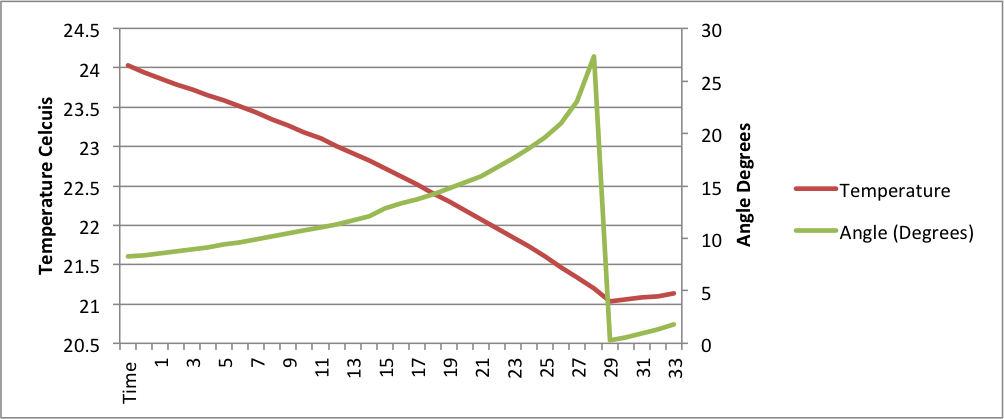}
\caption{Temperature change and angle computation to meet target temperature}
\label{figy}
\end{figure}
The data collected was used to compile Figure \ref{figy} which shows how the angle is adjusted in order to meet the target temperature. It also shows how the temperature changes as a result of a change in the shed angle. Initially the temperature is set to 24 Celsius and the user chooses 21 Celsius as target that should be achieved within 30 minutes. Figure \ref{figy} shows that the target is achieved within the required time. As a result, the user's comfort is increased because the target is achieved. 
\section{Conclusion and future work}
Our studies show that there is a potential for energy optimization in buildings and the central plant's machinery. To optimize the energy in the central plant, there is a need to monitor the efficiency profile of every machine, how one's efficiency affects the others and how this efficiency profile changes through time. therefore machine learning  is the appropriate approach in this changing environment. \\
The buildings were shown to be subject to optimization that will minimize the energy consumption and also increase the users' comfort. The airflow system simulation showed that there is a potential for energy efficiency. As a future work, this system will be tested in a real-world environment and serve as an infrastructure for sensing and control. In addition, the current model does not take into account real world parameters such as pressure of the room, the effect of one room air needs on the other rooms. These will be taken into consideration to make the model more robust to these changes.

\section*{Acknowledgment}
We would like to express our very great appreciation to Demond Williams for his valuable recommendations as well as his help in conducting this study. We would also like to express deep gratitude for Tao LV who has been part of this project, who has helped in the understanding of the system. We would also like to thank the Central plant operators who have spent with our group time to explain to us the the structure of the plant as well as to helping us conduct many experiments. We would like to thank the Central Plant in the University of Houston for giving us the chance to work with them and carry our project with their valuable assistance. We would like to also thank the students from College of Technology in University of Houston who spent long hours contributing to this project by helping digitizing the data.
\bibliographystyle{IEEEtran} 
\bibliography{IEEEabrv,refs} 

\begin{thebibliography}{10}
\providecommand{\url}[1]{#1}
\csname url@samestyle\endcsname
\providecommand{\newblock}{\relax}
\providecommand{\bibinfo}[2]{#2}
\providecommand{\BIBentrySTDinterwordspacing}{\spaceskip=0pt\relax}
\providecommand{\BIBentryALTinterwordstretchfactor}{4}
\providecommand{\BIBentryALTinterwordspacing}{\spaceskip=\fontdimen2\font plus
\BIBentryALTinterwordstretchfactor\fontdimen3\font minus
  \fontdimen4\font\relax}
\providecommand{\BIBforeignlanguage}[2]{{%
\expandafter\ifx\csname l@#1\endcsname\relax
\typeout{** WARNING: IEEEtran.bst: No hyphenation pattern has been}%
\typeout{** loaded for the language `#1'. Using the pattern for}%
\typeout{** the default language instead.}%
\else
\language=\csname l@#1\endcsname
\fi
#2}}
\providecommand{\BIBdecl}{\relax}
\BIBdecl

\bibitem{ref1}
\BIBentryALTinterwordspacing
D.~of~Energy. (2013) Building energy data book. [Online]. Available:
  \url{http://buildingsdatabook.eren.doe.gov}
\BIBentrySTDinterwordspacing

\bibitem{ref7}
D.~Adriano, A.~Page, A.~Elseewi, A.~Chang, and I.~Straughan, ``Utilization and
  disposal of fly ash and other coal residues in terrestrial ecosystems: A
  review,'' \emph{Journal of Environmental Quality}, vol.~9, no.~3, pp.
  333--344, 1980.

\bibitem{ref8}
H.~Farhangi, ``The path of the smart grid,'' \emph{Power and Energy Magazine,
  IEEE}, vol.~8, no.~1, pp. 18--28, 2010.

\bibitem{ref9}
R.~E. Brown, ``Impact of smart grid on distribution system design,'' in
  \emph{Power and Energy Society General Meeting-Conversion and Delivery of
  Electrical Energy in the 21st Century, 2008 IEEE}.\hskip 1em plus 0.5em minus
  0.4em\relax IEEE, 2008, pp. 1--4.

\bibitem{ref10}
A.-H. Mohsenian-Rad, V.~W. Wong, J.~Jatskevich, R.~Schober, and A.~Leon-Garcia,
  ``Autonomous demand-side management based on game-theoretic energy
  consumption scheduling for the future smart grid,'' \emph{Smart Grid, IEEE
  Transactions on}, vol.~1, no.~3, pp. 320--331, 2010.

\bibitem{ref11}
Z.~M. Fadlullah, M.~M. Fouda, N.~Kato, A.~Takeuchi, N.~Iwasaki, and Y.~Nozaki,
  ``Toward intelligent machine-to-machine communications in smart grid,''
  \emph{Communications Magazine, IEEE}, vol.~49, no.~4, pp. 60--65, 2011.

\bibitem{ref12}
W.~Kirsner, ``Chilled water plant design,'' \emph{HEATING PIPING AND AIR
  CONDITIONING-CHICAGO-}, vol.~68, pp. 73--80, 1996.

\bibitem{ref13}
W.~J. Fisk, ``Health and productivity gains from better indoor environments and
  their relationship with building energy efficiency,'' \emph{Annual Review of
  Energy and the Environment}, vol.~25, no.~1, pp. 537--566, 2000.

\bibitem{ref14}
N.-C. Yang and T.-H. Chen, ``Assessment of loss factor approach to energy loss
  evaluation for branch circuits or feeders of a dwelling unit or building,''
  \emph{Energy and Buildings}, vol.~48, pp. 91--96, 2012.

\bibitem{ref15}
S.~Mamidi, Y.-H. Chang, and R.~Maheswaran, ``Improving building energy
  efficiency with a network of sensing, learning and prediction agents,'' in
  \emph{Proceedings of the 11th International Conference on Autonomous Agents
  and Multiagent Systems-Volume 1}.\hskip 1em plus 0.5em minus 0.4em\relax
  International Foundation for Autonomous Agents and Multiagent Systems, 2012,
  pp. 45--52.

\bibitem{ref16}
R.~E. Edwards, J.~New, and L.~E. Parker, ``Predicting future hourly residential
  electrical consumption: A machine learning case study,'' \emph{Energy and
  Buildings}, vol.~49, pp. 591--603, 2012.

\bibitem{ref3}
M.~Wallace, R.~McBride, S.~Aumi, P.~Mhaskar, J.~House, and T.~Salsbury,
  ``Energy efficient model predictive building temperature control,''
  \emph{Chemical Engineering Science}, vol.~69, no.~1, pp. 45--58, 2012.

\bibitem{ref4}
H.~Jouhara and R.~Meskimmon, ``Experimental investigation of wraparound loop
  heat pipe heat exchanger used in energy efficient air handling units,''
  \emph{Energy}, vol.~35, no.~12, pp. 4592--4599, 2010.

\bibitem{Java7Spec}
\BIBentryALTinterwordspacing
J.~Gosling, B.~Joy, G.~Steele, G.~Bracha, and A.~Buckley, \emph{The Java
  Language Specification}, java se 7 edition~ed., California, USA, February
  2012. [Online]. Available:
  \url{http://docs.oracle.com/javase/specs/jls/se7/jls7.pdf}
\BIBentrySTDinterwordspacing

\end{thebibliography}

\end{document}